\documentclass[10pt,journal,compsoc]{IEEEtran}

\ifCLASSOPTIONcompsoc
  \usepackage[nocompress]{cite}
\else
  \usepackage{cite}
\fi
\ifCLASSINFOpdf
   \usepackage[pdftex]{graphicx}
   \graphicspath{{../pdf/}{../jpeg/}}
  \DeclareGraphicsExtensions{.pdf,.jpeg,.png}
\else
   \usepackage[dvips]{graphicx}
  \graphicspath{{../eps/}}
  \DeclareGraphicsExtensions{.eps}
\fi
\usepackage{amsmath}
\usepackage{array}
\usepackage{mdwmath}
\usepackage{booktabs}
 \usepackage{dcolumn}
 \usepackage{enumerate}
 \usepackage{colortbl}
 \definecolor{mygray}{gray}{.9}

\usepackage{url}

\newtheorem{theorem}{Theorem}
\newtheorem{corollary}{Corollary}

\begin{document}
%

\title{The Node-Similarity Distribution of Complex Networks and Its Applications in Link Prediction}
%
%
%

\author{Cunlai~Pu,~
        Jie~Li,~Jian~Wang,~\IEEEmembership{Member,~IEEE} and Tony Q. S. Quek~\IEEEmembership{Fellow,~IEEE} 
\thanks{
C. Pu is with the School of Computer Science and Engineering, Nanjing University of Science and Technology, Nanjing, 210094, China.  E-mail: pucunlai@njust.edu.cn. }
\thanks{J. Li was with the School of Computer Science and Engineering, Nanjing University of Science and Technology, Nanjing, 210094, China. He is currently with University of Science and Technology of China, Hefei 230026, China.  E-mail: jieli.holy@gmail.com. }
\thanks{J. Wang is with the School of Data Science, Fudan University, Shanghai, 200433, China. E-mail: jian\_wang@fudan.edu.cn (Corresponding author). }
\thanks{T. Q. S. Quek is with the Information Systems Technology and Design Pillar, Singapore University of Technology and Design, Singapore.  E-mail: tonyquek@sutd.edu.sg.}
}

\maketitle

\begin{abstract}
Over the years, quantifying the similarity of nodes has been a hot topic in complex networks, yet little has been known about the distributions of node-similarity. In this paper, we consider a typical measure of node-similarity called the common neighbor based similarity (CNS). By means of the generating function, we propose a general framework for calculating the CNS distributions of node sets in various complex networks. In particular, we show that for the Erd\"{o}s-R\'{e}nyi (ER)  random network, the CNS distribution of node sets of any particular size obeys the Poisson law. We also connect the node-similarity distribution to the link prediction problem. We found that the  performance of link prediction depends solely on the CNS distributions of the connected and unconnected node pairs in the network. Furthermore, we derive theoretical solutions of two key evaluation metrics in link prediction: i) precision and ii) area under the receiver operating characteristic curve (AUC). We show that for any link prediction method, if the similarity distributions of the connected and unconnected node pairs are identical, the AUC will be $0.5$. The theoretical solutions are elegant  alternatives of the  traditional experimental evaluation methods with nevertheless much lower computational cost.
\end{abstract}


%
\IEEEpeerreviewmaketitle

\section{Introduction}
%
%
%
%
\IEEEPARstart{I}{n} the field of network science, great progress has been  achieved in understanding the fundamental properties of  complex networks \cite{vespignani2018twenty,barabasi2016}. As one of those properties, node-similarity has largely promoted the formation and evolution of complex networks \cite{Papadopoulos12}.
In general, the similarity among a set of nodes can be judged  from the dimension of node properties. For instance, in social networks, individuals are similar in the sense of sex, race, job, religion, etc~\cite{mollgaard16}. However, the data availability and reliability of various node properties cannot always be guaranteed.
 Recently,  much attention has shifted to the structural similarity, which quantifies the node-similarity based only on the network topological information~\cite{ren2018structure,zhang2018measure}. In this direction, one of the most significant findings~\cite{lv11}  is that  node-similarity can be reflected by the information of common neighbors. In particular, if two nodes have more common neighbors, they are more similar in general. Such phenomenon has been observed in numerous types of networks \cite{liben07}. More delicate definitions of node-similarity further consider the degree of common neighbors~\cite{adamic03,zhou09}, path between nodes~\cite{lu09,katz53}, community structure~\cite{daminelli,sound12}, etc.

Node-similarity has important applications in many network prediction problems~\cite{ren2018structure}. As a typical network prediction problem, link prediction \cite{liben07} is grounded on the empirical observation that similar entities are  likely to interact with each other. In link prediction, each unconnected node pair is assigned with a similarity score. Those with large scores  are  assumed to have  missing or future links with high possibility.
  To design link prediction methods, tools from  network science~\cite{lv11,martinez17} and machine learning~\cite{chen2016scaling,duan2017ensemble,xiao20183} have been employed. Among these, the graph-based  methods has recently gained considerable popularity for their computationally efficiency and competitive prediction accuracy.

Depending on how much the network topological information is used, the graph-based methods for link prediction can be grouped into three categories: \romannumeral1)  local methods, \romannumeral2) global methods and \romannumeral3) quasi-local methods. The local methods include indices such as Common Neighbor (CN)~\cite{liben07}, Adamic-Adar (AA)~\cite{adamic03}, Resource Allocation (RA)~\cite{zhou09}, Preferential Attachment (PA)~\cite{barabasi99}, and many derivatives~\cite{martinez17,tan2014link,zhu2015information}. While these local methods use  the neighborhood topology information only, they have demonstrated promising accuracy with relatively low computational cost, and thus can be scalable to large and dynamic networks. However,  this type of methods  neglect the possibility of link formation among distant nodes. On the other hand, the global methods, such as the Katz~\cite{katz53}, Negated Shortest Path (NSP)~\cite{liben05} and Path Entropy (PE)~\cite{xuz16}, utilize the whole network topological information, and thus can give similarity scores to all node pairs, yet their computational cost is often inhibitive for large networks. The quasi-local methods, such as Local Path (LP)~\cite{lu09}, and Local Random Walks (LRW)~\cite{liuw10}, aim at pursuing balance between the local and global methods. Methods in this category can be applied to the friendship recommendation in social networks~\cite{wangd}, personalized product recommendation in e-commerce~\cite{xief,fu2018link}, structure and function analysis of biological networks~\cite{barzelb,sulaimany}, etc.

Although the similarity of two nodes has been extensively studied in the literature, how to quantify the similarity among an arbitrary number of nodes remains open. To date, little has been known about the node-similarity distribution of complex networks and its  applications in network prediction problems.    An obvious conclusion, for instance,  is that  links in the networks of even node-similarity distribution are difficult to predict. The  exploration of similarity among arbitrary many nodes will favor the research of higher-order link prediction~\cite{benson2018simplicial,Zhangm18}. In this paper, we would like to answer the following two fundamental questions: 
\begin{enumerate}[i)]
\item What are the node-similarity distributions of complex networks?
\item How to apply them to solve the link prediction problem?
\end{enumerate} 
We follow the definition of CN-based node-similarity. Assume that the similarity of a bunch of nodes equals the number of common neighbors these nodes share in the network, which is referred to as the {\it common neighbor based similarity (CNS)} hereafter. For the sake of generality, we also assume that the CNS of a single node is equal to its degree.

 Our main contributions are twofold.
 \begin{enumerate}[i)]
  \item Based on the generating function, we formalize the calculation of CNS distribution, and also investigate the CNS distributions for various types of complex networks, including regular networks, random networks, small-world networks, scale-free networks, and real-word networks. In particular, we found that in an ER random network, the CNS distribution of node sets of any given size follows the Poisson law.
   \item We  present theoretical solutions for two mainstream metrics of the link prediction accuracy: i)  the area under the
receiver operating characteristic curve (AUC) and ii)~precision \cite{lv11}, based on the node-similarity distributions of the connected and disconnected node pairs. We prove that for any link prediction method, if the similarity score distributions of connected and disconnected node pairs in a network are identical, the AUC will be~$0.5$.
 \end{enumerate}

The rest of the paper is organized as follows. In Section~\uppercase\expandafter{\romannumeral2}, we first present the calculation framework  of CNS distribution, and then explore the CNS distributions of various  network models and real-world networks. In Section~\uppercase\expandafter{\romannumeral3}, we provide the application of CNS distribution to the performance evaluation of  link prediction. Conclusion remarks are given in Section~\uppercase\expandafter{\romannumeral4}. 
\section{Common neighbor based similarity distribution}

Let $\mathcal{G}(\mathcal{V},\mathcal{E})$ be a graph, where $\mathcal{V}$ is the node set and $\mathcal{E}$ is the link set. The number of nodes in $\mathcal{V}$ is $N=|\mathcal{V}|$. Let $\mathbf{A}$ be the adjacent matrix of $\mathcal{G}$, in which $A_{ij}=1$ if nodes $i$  and $j$ are connected with a link; otherwise, $A_{ij}=0$. We have $$\sum_{i,j}A_{ij}=N\langle k \rangle,$$ where $\langle k \rangle$ is the average node degree. Also, let $V_q=\{v_1,v_2,\dots,v_q\} $ be a node set, where  $v_i\in \mathcal{V}$ and $q \in [1,N]$ is the size of $V_q$. Note that the node order is irrelevant in the node set. Then, the CNS of $V_q$ can be given~by
\begin{eqnarray}
\Theta(V_q)=\sum_{t\in \mathcal{V}-V_q}\delta(A_{v_1t}+A_{v_2t}+\dots+A_{v_qt}-q),
\end{eqnarray}
where $\delta(x)=1$ if $x=0$; otherwise, $\delta(x)=0$. Specially,  we have $$\Theta{\textbf(}\{v_i\}{\textbf)}=k_{v_i},$$ where $k_{v_i}$ is the degree of node $v_i$ and $\Theta(V_N)=0$.

Given an arbitrary node pair $(i,j)$, the probability that nodes $i$ and $j$ are connected with a link  is denoted by $\Gamma_{ij}$. Then, $\Gamma_{ij}=\Gamma_{ji}$. The probability that all  nodes in $V_q$ are connected with node $t$ is
\begin{equation}
\Gamma_{V_q\rightarrow t}=\prod_{v_i \in V_q}\Gamma_{v_it},
\end{equation}
and thus the probability that not all the nodes in $V_q$ are connected with node $t$ is $$\Gamma_{V_q\uparrow t}=1-\Gamma_{V_q\rightarrow t}.$$

With the above two probabilities, we can obtain the generating function $P[\Theta(V_q)=w]$~\cite{newmanmej01} of the probability distribution for the event that
the node set $V_q$ has $w$ common neighbor nodes, which is 
\begin{eqnarray}
G_{\Theta(V_q)}(x)&=& \nonumber \prod_{t\in \mathcal{V}-V_q}(\Gamma_{V_q\uparrow t}+\Gamma_{V_q\rightarrow t}x)\\
&=&\prod_{t\in \mathcal{V}-V_q}(1-\prod_{v_i \in V_q}\Gamma_{v_it}+x\prod_{v_i \in V_q}\Gamma_{v_it} ).~~~
\end{eqnarray}
According to the property of generating function, we have
\begin{eqnarray}
P[\Theta(V_q)=w]=\frac{G^{(w)}_{\Theta(V_q)}(0)}{w!}.
\end{eqnarray}
Furthermore, by the definition of probability distribution, the CNS distribution of node sets of size $q$ ($\geq 1$)  can be  given by
\begin{eqnarray}
P \left(\Theta^q=w \right) \nonumber &=&\frac{q!\sum_{V_q \subset \mathcal{V}}P \left[\Theta(V_q)=w \right]}{N(N-1)\cdots(N-q+1)}\\
&=&\frac{q!\sum_{V_q \subset \mathcal{V}}G_{\Theta(V_q)}^{(w)}(0)}{w!N(N-1)\cdots(N-q+1)}.
\end{eqnarray}

Thus far, we have presented a general CNS distribution calculation framework for the node sets of arbitrary size. Note that  node degree is a special case of CNS, where the node set contains only one node. Thus, the node degree distribution can  be  obtained through equations (2)--(5).
In  applications, we usually consider the CNS of  node sets with a specific size. For example, in the link prediction problem, the CNS of node sets of only two nodes are concerned. By setting $q=2$ to equation~(5), we can obtain the  CNS distribution  $P_a(\Theta^2=w)$ of all node sets containing two nodes.. We can also calculate the CNS distributions $P_c(\Theta^2=w)$ of connected node pairs and $P_d(\Theta^2=w)$ disconnected node pairs. In particular, the three distributions satisfy
\begin{equation}
\begin{cases}
 P_a(\Theta^2=w)= P_c(\Theta^2=w)\chi_c+P_d(\Theta^2=w)\chi_d,\\
\chi_c = \frac{\langle k \rangle}{N-1}=1-\chi_d.
\end{cases}
\end{equation}

In the following, we  shall explore the CNS distributions of different types of complex networks,  including various model networks and real-world networks.

\subsection{\label{sec:level2}Regular Ring Lattice}
We consider a regular ring lattice (RRL)~\cite{newman2010networks}, in which all the nodes are located on a ring. Node $i$ ($i=1, 2,\dots,N$) has $ m$ neighbors in each of the two sides, and thus the average node degree $\langle k \rangle$ is $2m$. For an arbitrary node pair $(i,j)$ in the RRL, the connection probability is $$\Gamma_{ij}=\delta(|i-j|\leq m).$$ Then, through equations (2)--(5) we can obtain the CNS distribution of regular networks.
 \begin{figure}
\centering
\includegraphics[width=3.5in]{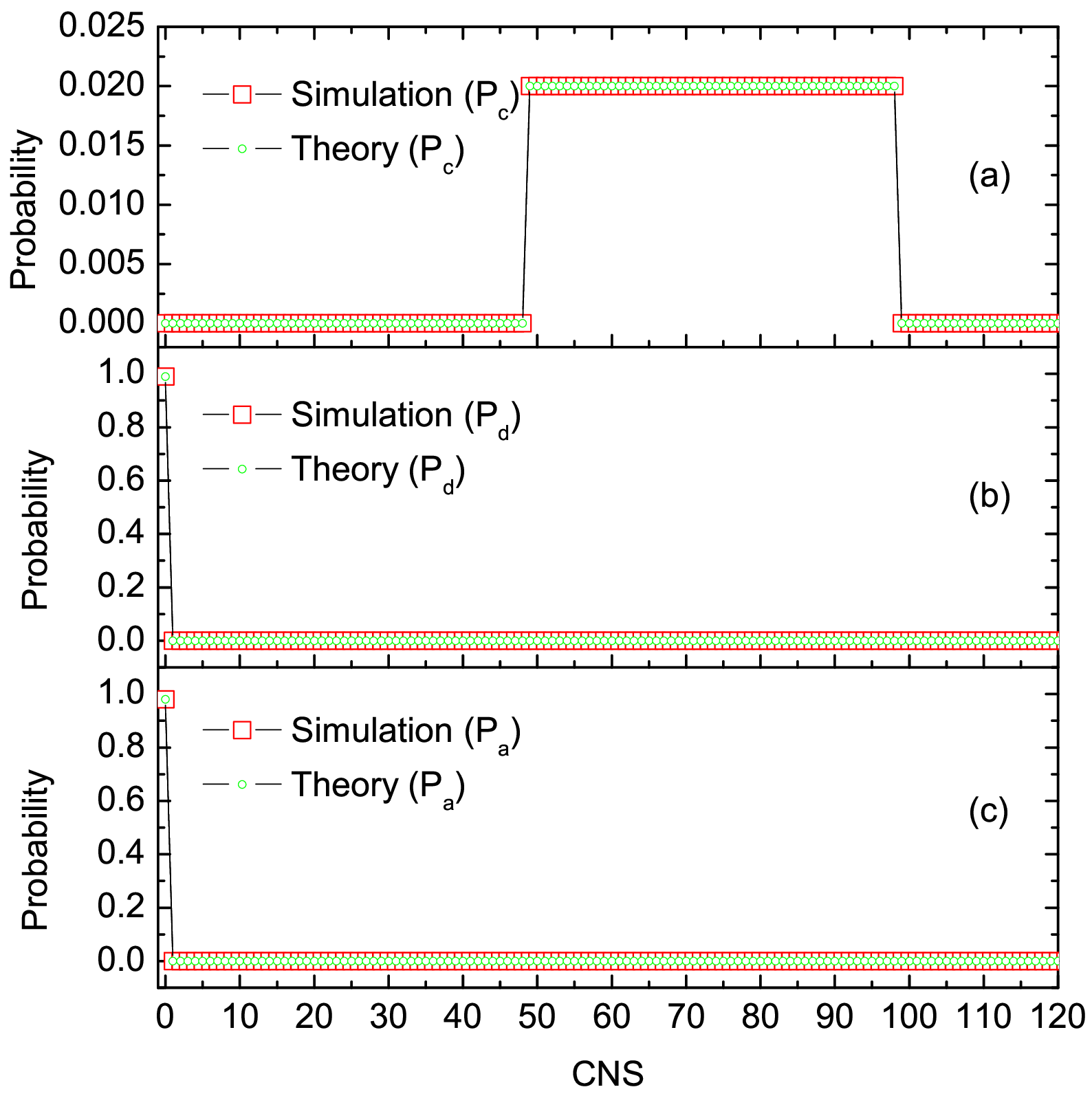}
\caption{The CNS distributions of (a) connected, (b) disconnected, and (c) all node pairs in a regular network. }
\end{figure}

Figure 1 shows the  CNS distribution of node pairs  in a  regular network with $N=1,000$ and $m=50$. We observe that the simulation and analytical results  are well matched. Specifically, in Figure  1(a), the CNS is bounded within $[m-1,2m-2]$ and $P_c(w)=1/m$ for connected node pairs. Figure 1(b) shows that for disconnected node pairs, $$P_d(0)=\frac{N-1-4m}{N-1-2m}\approx 1.$$ Whereas for a wide range of CNS, $$P_d(w)= \frac{2}{N-1-2m} \approx 0, ~~w=1,2,\dots,m.$$  Since $\chi_c=2m/(N-1)\approx 0$, the CNS distribution of all node pairs is dominated by the disconnected node pairs, as shown in Figure 1(c).

\subsection{ The ER Network Model}
 In this section, we consider the Erd\"{o}s-R\'{e}nyi (ER)  random network model~\cite{erdos} of $N$ nodes. The connection probability of any node pair in the ER network is $$p^{\text{E}}=\frac{2n}{N(N-1)},$$ where $n$  is the expected number of links. The average node degree is $\langle k \rangle=2n/N\approx p^{\text{E}}N$.
    Then, we have the following result.
    
    \vspace{1mm}
    \begin{theorem}
  In the ER model, for an arbitrary node set $V_q$, the probability that the nodes in  $V_q$ have $w$ common neighbor nodes is
 \begin{equation}
 P\left[\Theta(V_q)=w \right]\approx e^{-\lambda}\frac{\lambda^w}{w!},
 \end{equation}
 where $\lambda=\langle k \rangle^qN^{1-q}$.
 \end{theorem}
\begin{IEEEproof}
By definition all nodes are equivalent in the ER model. For an arbitrary node pair $(i,j)$, the connection probability is $$\Gamma_{ij}=\frac{\langle k \rangle}{N-1}.$$ 
Substituting this into equation (3) yields
\begin{eqnarray}
G_{\Theta(V_q)}(x) \nonumber &=& \prod_{t\in \mathcal{V}-V_q}(1-\prod_{v_i \in V_q}\Gamma_{v_it}+x\prod_{v_i \in V_q}\Gamma_{v_it})\\
&=&\left[1-\frac{\langle k\rangle^q}{(N-1)^q}+\frac{\langle k \rangle^qx}{(N-1)^q}\right]^{N-q}.
\end{eqnarray}
By combining equations (4) and  (8), we have
\begin{eqnarray}
P\left[\Theta(V_q)=w \right] & = & \frac{G^{(w)}_{\Theta(V_q)}(0)}{w!}  \nonumber \\
& = & \begin{pmatrix}{ N-q} \\ w \end{pmatrix} \left(\frac{\langle k \rangle^q}{(N-1)^q}\right)^w \nonumber \\
&& \times \left(1-\frac{\langle k \rangle^q}{(N-1)^q}\right)^{N-q-w},
\end{eqnarray}
which is a binomial distribution. Assuming that $N \to \infty$, the binomial distribution is equivalent to the Poisson distribution, and so we have
\begin{align}
P\left[\Theta(V_q)=w \right]
\approx e^{-\lambda}\frac{\lambda^w}{w!},
\end{align}
where $\lambda=\langle k \rangle^qN^{1-q}$.
\end{IEEEproof}

\vspace{1mm}
 \begin{corollary}{}
  In an ER random network, the CNS distribution of node sets of size $q$ approximately follows the Poisson law,
  \begin{equation}
  P(\Theta^{q}=w) \approx e^{-\lambda}\frac{\lambda^w}{w!},
  \end{equation}
  where $\lambda=\langle k \rangle^qN^{1-q}$.
 \end{corollary}
 \begin{IEEEproof}
Immediate from  (5) and  (10).
\end{IEEEproof}


\vspace{2mm}
Corollary 1 indicates that the CNS distribution of node sets in the ER random network becomes more uneven when the average node degree increases, or when the network size (or node set size) decreases.

\vspace{1mm}
 \begin{corollary}{}
   Given an arbitrary set  $\Lambda=\{V_q\}$, in which the element $V_q$ is  a set of $q$ nodes selected from an ER network, if the cardinality of the set $|\Lambda|$ is  sufficiently large,  the CNS distribution of elements  in set  $\Lambda$ satisfies
\begin{equation}
P_{\Lambda}(\Theta^q=w)\approx P(\Theta^q=w).
\end{equation}
 \end{corollary}
\begin{IEEEproof}
Based on the definition of probability distribution,
\begin{eqnarray}
P_{\Lambda}(\Theta^q=w)= \frac{\sum_{\Lambda}\delta(\Theta(V_q)-w)}{|\Lambda|}.
\end{eqnarray}
Since $|\Lambda|$ is sufficiently large, we obtain
\begin{eqnarray}
 \frac{\sum_{\Lambda}\delta(\Theta(V_q)-w)}{|\Lambda|}\approx \frac{\sum_{\Lambda}P(\Theta(V_q)=w)}{|\Lambda|}.
 \end{eqnarray}
 Plugging equation (10) into equation (14), we obtain
 \begin{eqnarray}
 P_{\Lambda}(\Theta^q=w)\nonumber & \approx &\frac{\sum_{\Lambda}e^{-\lambda}\frac{\lambda^w}{w!}}{|\Lambda|}\\
&=& e^{-\lambda}\frac{\lambda^w}{w!}=P(\Theta^q=w).
\end{eqnarray}
\end{IEEEproof}

According to Corollary 2, we have that in  sufficiently large ER networks, $$P_c(\Theta^2=w)=P_d(\Theta^2=w)=P_a(\Theta^2=w)=e^{-\lambda}\frac{\lambda^w}{w!},$$
where $\lambda=\langle k \rangle^2/N$. In fact, this can also been  obtained  from the simulation results. As shown in Figure 2, the parameter of the ER random network is  $N=10,000$ and $\langle k \rangle=500$, which well matches the theoretical prediction.
 \begin{figure}
\centering
\hspace{-2mm}\includegraphics[width=3.6in ]{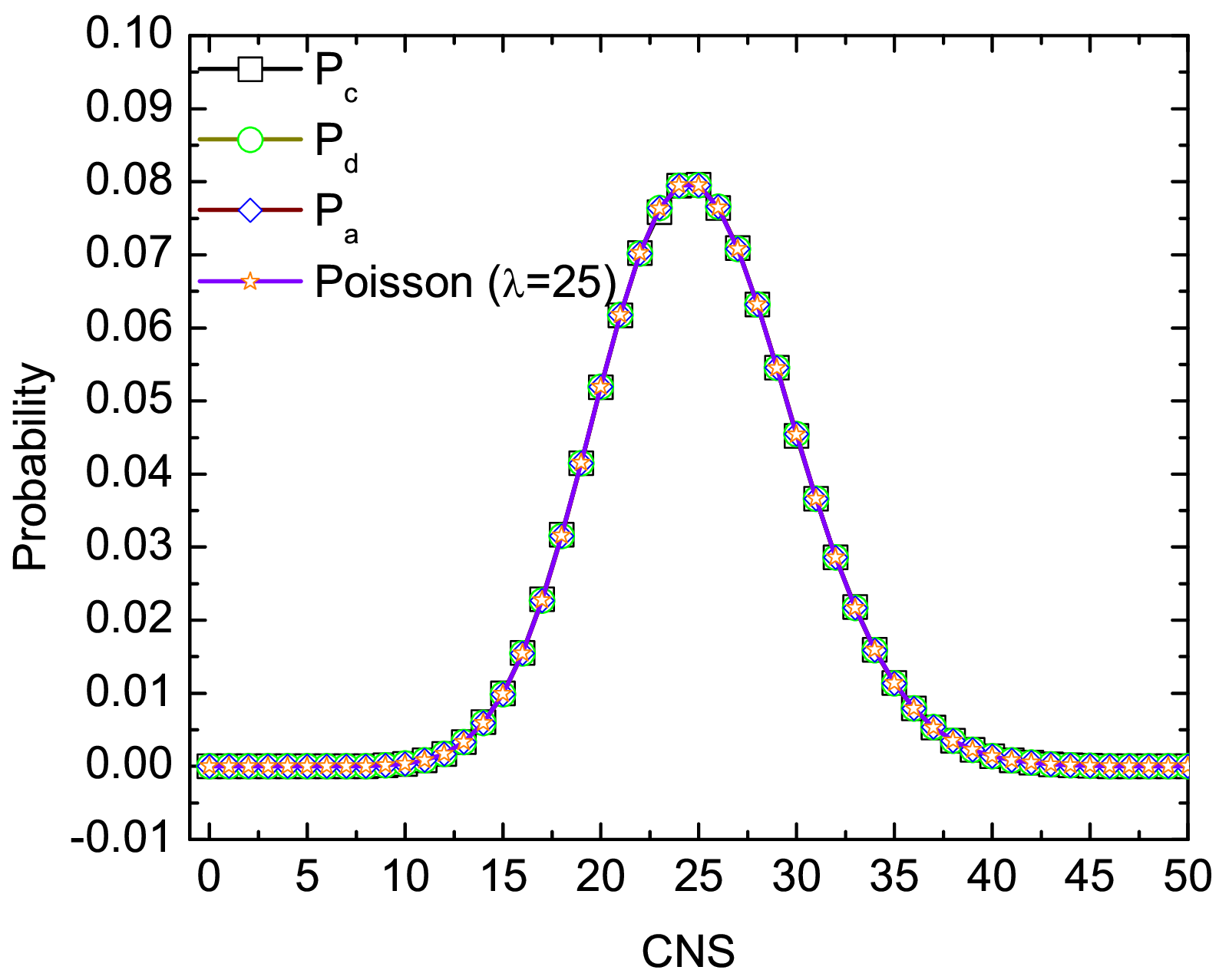}
\caption{The CNS distributions of   connected,  disconnected, and  all node pairs in an ER random network. }
\end{figure}
\subsection{The Small-World Network Model}
 We derive the CNS distribution of the  Watts and Strogatz~(WS)  model \cite{small-world1}, which is one of the most typical models that generate small-world networks.  In the WS model, the initial network is a regular ring lattice of $N$ nodes. Each node is connected to its  $2m$ nearest neighbors. Each link is randomly rewired with probability $p^{\text{WS}}$, where multiple links and self-connections are not allowed. In this case, the expected number of rewired links is $p^{\text{WS}} mN$. The average node degree equals $2m$,  which is unchanged before and after the random rewiring process.
Thus, the connection probability of a node pair $(i,j)$ can be given by
\begin{equation}
\Gamma_{ij}\hspace{-0.1cm}=\hspace{-0.1cm}\delta(|i\hspace{-0.05cm}-\hspace{-0.05cm}j|\leq m)(1\hspace{-0.05cm}-\hspace{-0.05cm}p^{\text{WS}})\hspace{-0.05cm}+\hspace{-0.05cm}\frac{2p^{\text{WS}}mN\delta(|i\hspace{-0.1cm}-\hspace{-0.1cm}j|\hspace{-0.1cm}>\hspace{-0.1cm}m)}{N(N\hspace{-0.1cm}-\hspace{-0.1cm}1)\hspace{-0.1cm}-\hspace{-0.1cm}2mN}.
\end{equation}
Substituting equation (16) into equations (2)--(5), we obtain the CNS distribution of the WS model.
 \begin{figure}
\centering
\includegraphics[width=3.5in ]{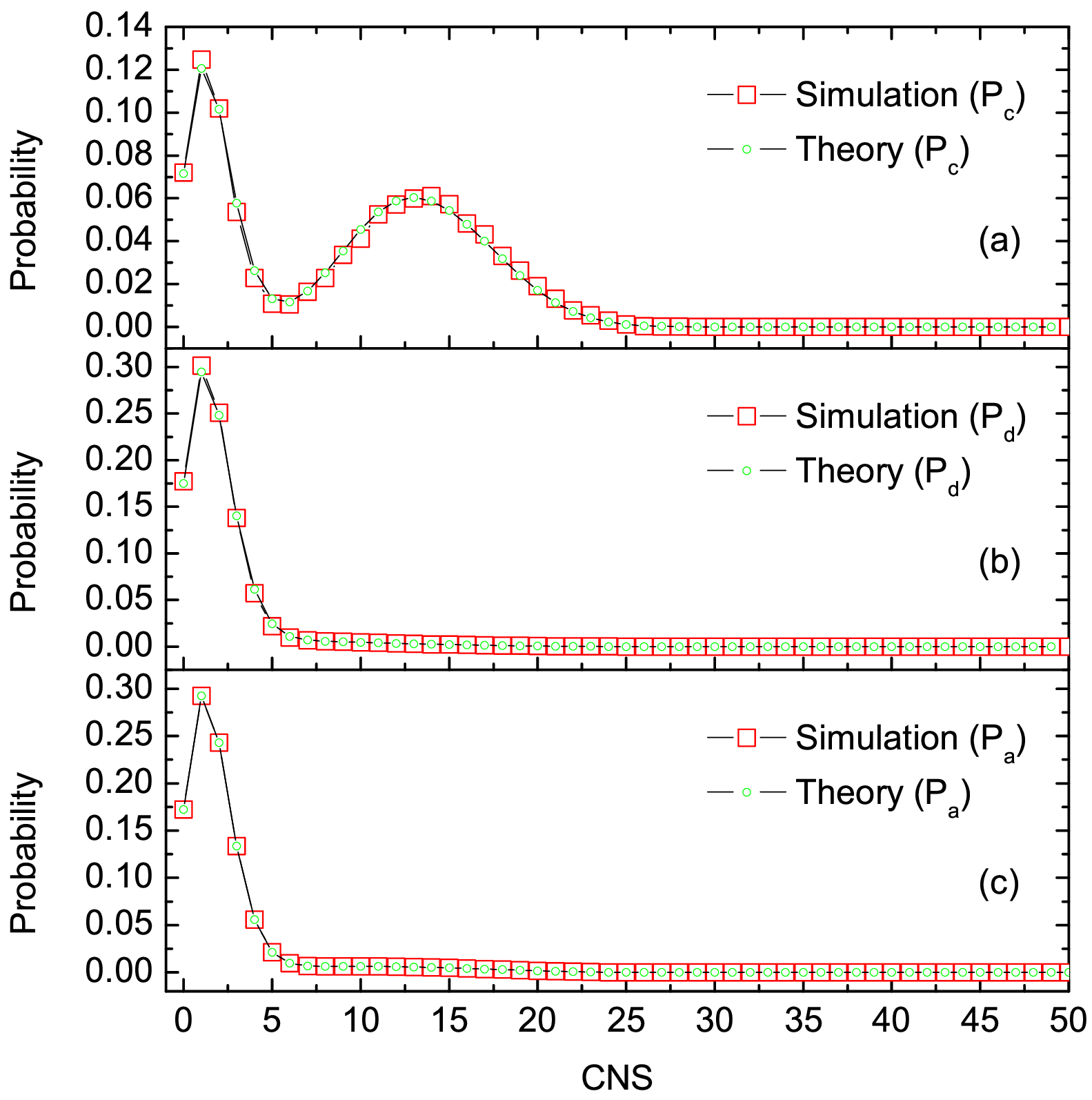}
\caption{The CNS distributions of   (a) connected, (b) disconnected, and (c) all node pairs  in a WS small-world network. }
\end{figure}

Figure 3 depicts the CNS distribution of node pairs in a WS small-world network  with $N=1,000$, $ m=25$ and $p^{\text{WS}}=0.4$. One can observe that the simulation results are consistent with the analytical results. Indeed, in Figure~3(a), the CNS for the connected node pairs clearly exhibits a bimodal distribution. Whereas in Figures 3(b) and (c),  the CNS  distributions for the disconnected node pairs and all node pairs approach the Possion law with a long tail.
\subsection{The Unified Ring Model}
The RRL is a very basic model. By deleting, adding and rewiring the links among nodes with certain probability, RRL can evolve  into many other models, e.g., the WS, NW and ER models. Particularly,  deleting each link  from a RRL with probability $p^{\text{M}}$ yields a  new model called modified ring lattice (MRL).
The average node degree of  MRL  is $$\langle k \rangle=2(1-p^{\text{M}}) m.$$ The connection probability between an arbitrary node pair $(i,j)$ in the MRL can be given by $$\Gamma_{ij}=\delta(|i-j|\leq m)(1-p^{\text{M}}).$$

We further provide a general  model called the unified ring model to unify the ring-based models,  including the RRL, MRL, WS, NW, and ER models. In the unified  model,  there is initially an empty  ring (without links) of $N$ nodes located  with even intervals. The nodes are labeled    clockwise from  $1$ to $N$ with a designated starting node. We then connect each node pair with probability $\eta$ if the corresponding node label distance is no greater than $m$; otherwise with probability $\alpha$.

The  connection probability of a node pair $ (i,j)$ in the unified ring model can be  expressed as
\begin{equation}
\Gamma_{ij}=\delta(|i-j|\leq m)\eta+\delta(|i-j|>m)\alpha.
\end{equation}
Substituting equation (17) into equations (2)--(5), we obtain the CNS distribution of  the unified ring model.

We specially consider a node set of two nodes, $V_2=\{i,j\}$,  in the unified ring model.
   For the node pair $(i,j)$, any other node $t$ should be in one of the three types, depending on the label distance $|t-i|$ and $|t-j|$.
   \begin{itemize}
  \item The first type  satisfies $|t-i|\leq m$ and $|t-j|\leq m$.  Assume that the number of nodes of  this type is $s_{ij}$. Then, we have
       \begin{equation}
       ~~~~s_{ij} =
        \begin{cases}
        2m-1-|i-j|,&  \text{if} ~0<|i-j| \leq m-1, \\
         2m-|i-j|, & \text{if} ~m \leq |i-j| \leq 2m, \\
         0, &\text{otherwise}.
        \end{cases}
    \end{equation}
 For any node $t$ of the first type, we have $$\Gamma_{\{i,j\}\rightarrow t}= \eta^2.$$
  \item The second type  satisfies either of  the following conditions: \romannumeral1) $|t-i|\leq m$ and $ |t-j|> m$ and \romannumeral2) $|t-i|> m$ and $|t-j| \leq m$ with $\Gamma_{\{i,j\}\rightarrow t}= \eta\alpha$, and the number of nodes of  this type  is   $4m-2s_{ij}$.
       \item The last type  satisfies $|t-i|> m$ and $|t-j|> m$ with $\Gamma_{\{i,j\}\rightarrow t}= \alpha^2$. In this case,  the number of nodes is $N-2-4m+s_{ij}$.
           \end{itemize}
Thus, by equation (3) we obtain  the generating function of the CNS  distribution  of  node set $ \{i,j\}$ as:
\begin{align}
G_{ij}(x)=\nonumber&[1-\eta^2+\eta^2x]^{s_{ij}}(1-\eta \alpha+\eta \alpha x)^{4m-2s_{ij}}\\
&\times(1-\alpha^2+\alpha^2x)^{N-2-4m+s_{ij}}.
\end{align}

The ring based models, including the RRL, MRL, ER, WS, and NW models, are  special cases of the unified ring model.
The corresponding $\eta$, $\alpha$, and  generating function for each of these models are summarized  in Table 1.

\begin{table*}
\caption{Connection probability $\eta$, $\alpha$ and  generating function  $G_{ij}(x)$ for different network models.}
\scriptsize
\begin{tabular*}{\textwidth}{@{}l*{15}{@{\extracolsep{0pt plus
12pt}}l}} \toprule
Model&$\eta$&$\alpha$&$G_{ij}(x)$\\
\midrule
RRL&1&0&$x^{s_{ij}}$\\
MRL&$1-p^M$&0&$(1-\eta^2+\eta^2x)^{s_{ij}}$\\
ER &$p^E$&$p^E$&$(1-\alpha^2+\alpha^2x)^{N-2}$ \\
WS&1-$p^{WS}$&$2mp^{WS}/(N-1-2m)$&$(1-\eta^2+\eta^2x)^{s_{ij}}(1-\eta\alpha+\eta\alpha x)^{4m-2s_{ij}}(1-\alpha^2+\alpha^2x)^{N-2-4m+s_{ij}}$ \\
NW&1&$2mp^{NW}/(N-1-2m)$&$x^{s_{ij}}(1-\alpha+\alpha x)^{4m-2s_{ij}}(1-\alpha^2+\alpha^2x)^{N-2-4m+s_{ij}}$ \\ \bottomrule
\end{tabular*}
\end{table*}
\normalsize

\subsection{The Scale-Free Network Model}
We now consider the  Barab\'{a}si-Albert (BA) model \cite{barabasi99},  which is one of the most typical models for generating scale-free networks. In the BA model,  there are initially $m_0$ fully connected nodes labeled as $1,2,\cdots,m_0$. For each time a new node, labeled from $m_0+1$, is added into the network with $m~(\leq m_0)$ links connecting the existing nodes until the total number of nodes reaches $N$.  The probability that an existing node is selected for each link is proportional to its  degree. For instance, the probability that the new node $i$ selects node $j \in [1,i-1]$ is 
\begin{equation}
 p_j(i) = \frac{k_j(i)}{\sum_{t=1}^{i-1}k_t(i)},
 \end{equation} 
 where $k_j(i)$ is the degree of node $j$ when node $i$ is just added into the network.

 For the generated BA  network, the average node degree is $\langle k \rangle =2m$, and the node degree distribution obeys the power law with exponent three. When constructing the links of a new node, the overall times of node selection should be no less than $m$ because of  duplication. Assume that for node~$i$ the expected node selection times in the link construction process  is $T_i$. Then, the overall connection probability of nodes $i$ and  $j$ with $j<i$ satisfies
\begin{equation}
\begin{cases}\Gamma_{ij}=1-(1-p_j(i))^{T_i},\\
\sum_{j<i}\Gamma_{ij}=m.
\end{cases}
\end{equation}
 By equation (20) we obtain $\Gamma_{ij}$ and thus $\Gamma_{ji}$ (since they are equal). Thus, we update the node degree with $$k_j(i+1)=k_j(i)+\Gamma_{ji}.$$

  After obtaining the  connection probabilities of all node pairs, we substitute them into equations (2)--(5) to calculate the CNS distribution of the BA model.
 Figure 4 shows the CNS distribution of node pairs in a  BA scale-free network  with $N=1,000$, and $\langle k \rangle=50$, where the simulation results and analytical results  are consistent with each other. Clearly all the three CNS distributions, $P_c$, $P_d$ and $P_a$,  approximately follow the power law distributions with long tails, whose   parameters  are  given in the corresponding subfigures, respectively.
 \begin{figure}
\centering
\includegraphics[width=3.5in]{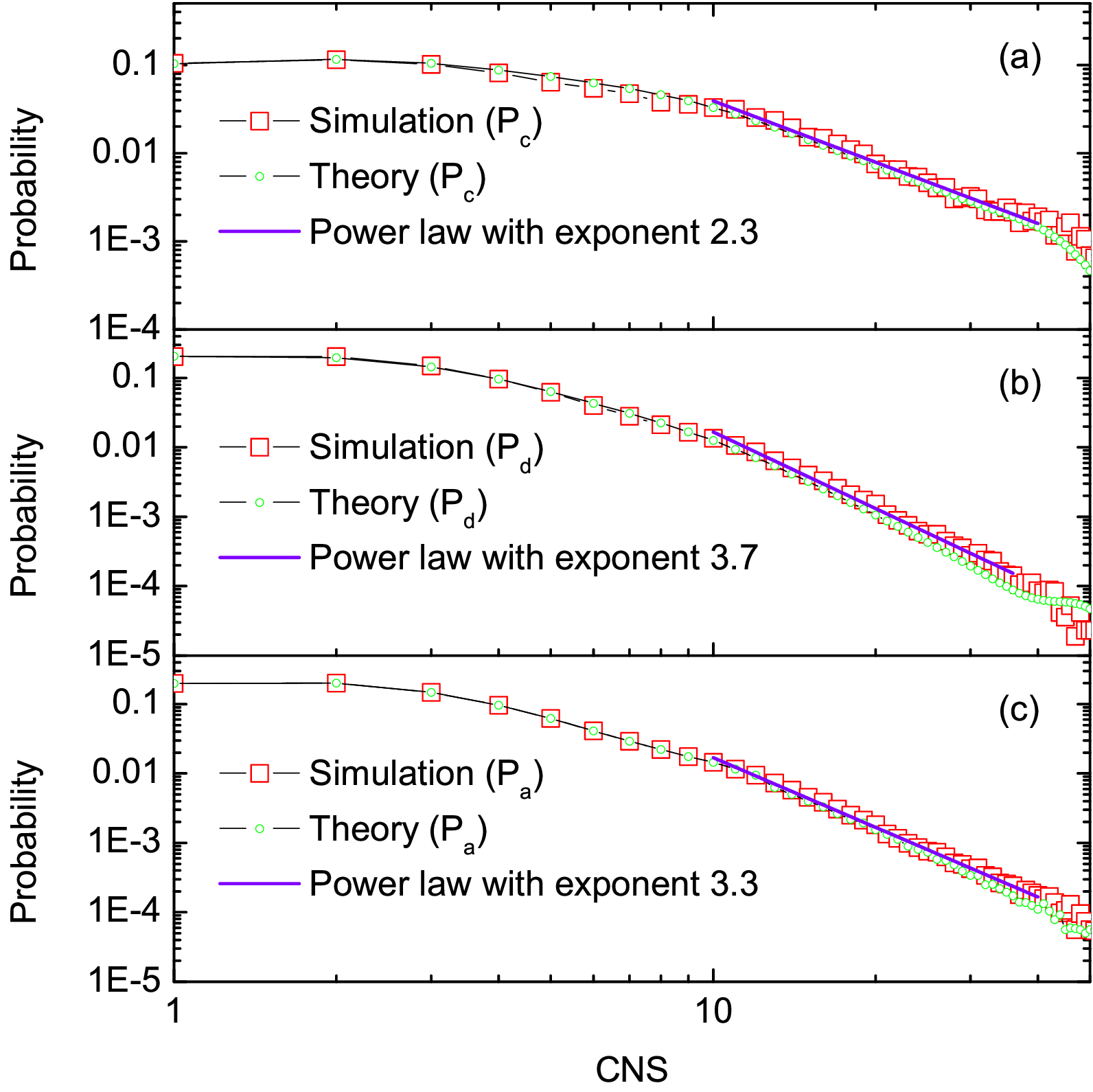}
\caption{The CNS distributions of  (a) connected, (b) disconnected, and (c) all node pairs in a BA scale-free network. }
\end{figure}
\subsection{Real-World Networks}

 We also consider  real-world networks and select some typical network data downloaded from the Stanford Network Analysis Platform (SNAP)~\cite{dataset1} and the Koblenz Network Collection (KONECT)~\cite{dataset2}, whose statistics are marked on Figure 5. All the network data are processed by ignoring the directions of  links and  deleting the multiple and self-connections. The CNS distributions of  connected node pairs in these real-world networks  are shown in Figure 5.  We can observe that for most of the real-world networks, the CNS distributions of  connected node pairs  approximately follow the power law with various  parameters. We can also observe that for some specific real-world networks, the CNS of connected node pairs  obeys   the  exponential law.
    \begin{figure*}
\centering
\includegraphics[width=7.3in]{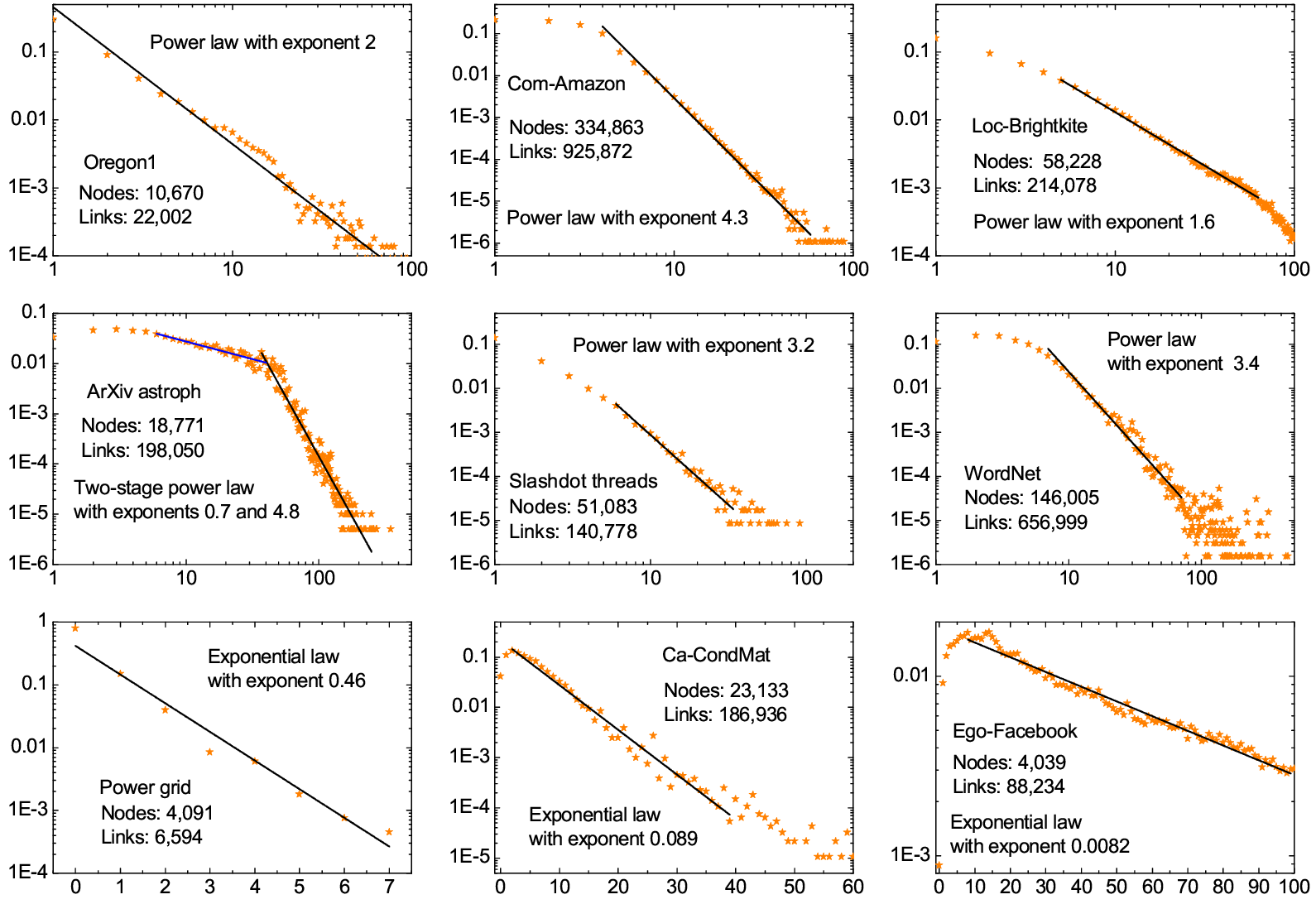}
\caption{The CNS distributions of   connected node pairs  in some real-world networks. $x$-axis: CNS, $y$-axis: Probability. }
\end{figure*}

\section{Application}

In this section, we give a direct application of node-similarity distribution to  the link prediction problem. Specifically, we first provide the framework to evaluate the prediction accuracy and introduce the definition of two traditional metrics: AUC and precision \cite{lv11}. Next, we derive the AUC  and precision based on the  similarity distributions of connected and disconnected node pairs. Then, we present a theorem of AUC based on  its theoretical solution. Finally, we provide some simulation results to further validate our theoretical findings.


\subsection{Prediction Accuracy Evaluation}
A common assumption in  link prediction  is that the connection probability of two nodes is  positively correlated with the degree of their  similarity.
To evaluate the prediction accuracy of a  similarity index, we  randomly select a small fraction (10\% or 20\%) of links from the original network as the test set, and the remaining network forms the training set.
  Then, we calculate the similarity scores of all  disconnected node pairs in the training set based on the given similarity index and  the network structure of training set.


Let us further divide   the  disconnected node pairs of the training set into two sets. The first set contains the node pairs whose links are moved into the test set. The second set consists of the node pairs that are not connected in the original network. Each time we compare the similarity scores of a node pair randomly selected  from the first set and another one randomly selected  from the second set. We carry out this comparison  $n$ times in total. If there are $n'$ times that the former is larger and $n''$ times that they are equal, we have
\begin{equation}
AUC=\frac{n'+0.5n''}{n}.
\end{equation}
From this definition, we see that a valid link prediction method should have an AUC value larger than 0.5.

Besides AUC, another typical metric of prediction accuracy is  precision, denoted by \textit{prec}.
To calculate it, we  rank the disconnected node pairs of the training set  (union of the first set and second set) with decreasing order of  similarity scores. If $ L'$ out of the top-$L$ node pairs belong to the first set,  then we have
\begin{equation}
\textit{prec}=\frac{L'}{L},
\end{equation}
where $L$ is usually set to be  the size of the test set.

\subsection{Theoretical Solutions of AUC and Precision}
We show that the prediction accuracy of the CN index is determined by the CNS distributions of connected and disconnected node pairs, i.e., $P_c$ and $P_d$, which can be extended to the other similarity-based link prediction methods.
Statistically, randomly removing a small percent of links does not significantly affect the CNS distribution of the network. Thus, we assume that the CNS distributions of connected and unconnected node pairs are approximately the same before and after the training set and test set division, which is the key precondition of the following calculation.

Assuming that the number of links in the test set is $\epsilon N \langle k \rangle/2$, $0<\epsilon<1$, the AUC can be calculated as
\begin{equation}
AUC = \sum^{N-2}_{x=0}P_c(w=x)\left[P_d(w<x)+\frac{P_d(w=x)}{2}\right].
\end{equation}

Moreover, we rank all the disconnected node pairs  of the training set,   union of the first and second sets defined in the above,  with decreasing order of CNS scores, and then consider the top-$L$ node pairs.
Let us denote by $\Phi_c(x)$, $\Phi_d(x)$, and $\Phi(x)$   the number of disconnected node pairs (with the CNS scores no less than $x$) belonging to the first, second, and  whole training set, respectively.
 Then, we have
\begin{equation}
\begin{cases}\Phi_c(x)=\sum^{N-2}_{w=x}\epsilon\frac{N\langle k \rangle}{2}P_c(w), \vspace{2mm} \\
\vspace{1mm}
\Phi_d(x)=\sum^{N-2}_{w=x} \frac{N(N-1-\langle k \rangle)}{2}P_d(w), \vspace{1mm} \\
\Phi(x)= \Phi_c(x)+ \Phi_d(x).
\end{cases}
\end{equation}
We obtain the CNS score $x_L$ of the $L$-th node pair by further  solving the following inequality:
\begin{equation}
\Phi(x+1)\geq L> \Phi(x).
\end{equation}
The precision can thus be  calculated as
\begin{eqnarray}
 \lefteqn{\textit{prec} =  \frac{\Phi_c(x_L + 1)}{L}} ~~~\nonumber \\
 &&+ \frac{N\langle k \rangle \epsilon P_c(x_L) [L-\Phi(x_L+1)]}{L N \langle k \rangle \epsilon P_c(x_L)+  LN(N - 1 - \langle k \rangle) P_d(x_L)}.~~~~~~
\end{eqnarray}
Loosely, we have
\begin{equation}
\textit{prec} \approx \frac{\Phi_c(x)}{\Phi(x)}.
\end{equation}

Through the above calculation, we obtain the theoretical solutions of the AUC and precision, which avoid  the actual training and test set division and   the associated computational cost.
It should be mentioned that although we only take the CN index as an example to present the theoretical calculation, the calculation framework is applicable to all the similarity-based link prediction methods. Note that for other similarity indices, the CNS distribution should be replaced by the corresponding node-similarity distributions in the calculation.

  Based on the theoretical solutions of the AUC and precision, we can explore the fundamental limit of the link predictability of complex networks.
      \begin{theorem}{}
      For any link prediction method, if the  similarity score distributions $P_c(w)$ and $P_d(w)$ of connected and disconnected node pairs in a  network are identical,   then the AUC of this prediction  method on this network equals $0.5$.
 \end{theorem}
\begin{IEEEproof}
Since $P_c(w)$ and $P_d(w)$ are identical, we assume that $P_c(w)=P_d(w)=P(w)$. Substituting this into equation~(23), we have
\begin{equation}
AUC = \sum^{N-2}_{x=0}P(w=x)\left[P(w<x)+\frac{P(w=x)}{2}\right].
\end{equation}
Note that
\begin{eqnarray}
\sum^{N-2}_{x=0}P(w=x)P(w<x)\nonumber \! \! \! \!&=& \! \! \!  \!\sum^{N-2}_{x=0}P(w=x)\sum^{N-2}_{y=0}P(w=y)\\
\nonumber  \!  \! \!\! &=& \!  \! \!\! \sum^{N-2}_{y=0}P(w=y)\sum^{N-2}_{x=y+1}P(w=x)\\
 \! \! \! \!&=& \! \! \! \! \sum^{N-2}_{x=0}P(w=x)P(w>x).
\end{eqnarray}
Thus, we have
\begin{align}
AUC+AUC=\nonumber&\sum^{N-2}_{x=0}P(w=x)\left[P(w<x)+\frac{P(w=x)}{2}\right]\\ \nonumber
&+\sum^{N-2}_{x=0}P(w=x)\left[P(w>x)+\frac{P(w=x)}{2}\right]\\ \nonumber
&=1.
\end{align}
The proof is complete.
\end{IEEEproof}

Theorem 2 is quite general, and a special case is the  ER random network. Based on Corollary 2, when the CN index is used on the ER random network, $P_c(w)=P_d(w)$. Then, according to Theorem 2, the AUC of the ER random network is 0.5.

\subsection{Simulation Results}
To further validate our theoretical solutions of AUC and precision, we conduct  experiments on link prediction over three real-world networks \cite{dataset3}, of which the statistics are provided in Table II. The self-connections and multiple links are deleted from the network data. The CN, RA, AA, LP, and Katz indices (see Appendix) are used here as the examples of link prediction method, where CN, RA, AA are local indices, LP is a quasi-local index, and Katz is a global index.    In the experiment, we perform 100 trials of the test  and training sets division according to the 10/90 rule. When calculating the AUC, the times of comparison  are $n=10^4$. For the precision, $L$ is equal to the size of the test set. The theoretical results are calculated by using equations~(23) and~(26).  For each index, we show both the experimental and theoretical results of AUC and precision in Table II, from which we  can see that these results are well matched.
\begin{table*}
\caption{Link prediction performance of three real-world networks.}
\scriptsize
\begin{tabular*}{\textwidth}{@{}l*{15}{@{\extracolsep{0pt plus
12pt}}l}} \toprule
Jazz (nodes: 198, links: 5484)\\ \hline
Performance&CN&RA&AA&LP&Katz\\
Experimental AUC&0.954&0.973&0.963&0.948&0.940\\
Theoretical AUC&0.958&0.974&0.965&0.950&0.947\\
Experimental precision&0.508&0.555&0.520&0.501&0.443\\
Theoretical precision&0.523&0.551&0.536&0.478&0.464\\ \hline
Usair (nodes: 1532, links: 2126)\\\hline
Performance&CN&RA&AA&LP&Katz\\
Experimental AUC&0.934&0.954&0.945&0.931&0.919\\
Theoretical AUC&0.941&0.960&0.953&0.955&0.951\\
Experimental precision&0.359&0.459&0.397&0.382&0.350\\
Theoretical precision&0.371&0.474&0.393&0.385&0.376\\\hline
Political blogs (nodes: 1222, links: 19021)\\\hline
Performance&CN&RA&AA&LP&Katz\\
Experimental AUC&0.918&0.923&0.923&0.927&0.924\\
Theoretical AUC&0.923&0.928&0.926&0.953&0.938\\
Experimental precision&0.170&0.147&0.169&0.173&0.172\\
Theoretical precision&0.175&0.153&0.173&0.199&0.183\\\bottomrule
\end{tabular*}
\end{table*}

In fact, we can even gain some insights of the prediction accuracy from the  similarity distributions of the connected and disconnected node pairs in the network without calculating the AUC and precision. Let us again take the CN index for example.  In Figure 6,  the curves of $P_c$  lie uniformly at the top right of those  of $P_d$ for all the three real-world networks,  which  confirms the common assumption of link prediction that node pairs with more common neighbors are more prone to be connected. The farther the two distributions are apart, which can be measured by the distance of their medians in Figure 6, the better the prediction performance will be.  This conclusion is obtained by combining the results of Table 2 and Figure 6, which can also be inferred from equation (28).
\begin{figure*}
\centering
\hspace{-3mm}\includegraphics[width=7.3in]{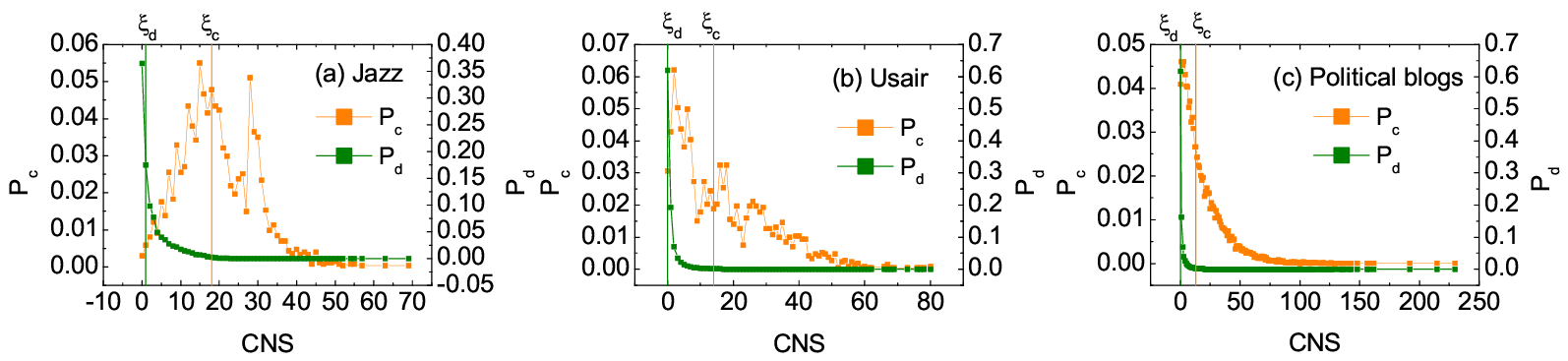}
\caption{CNS distributions of the connected (orange) and disconnected (green) node  pairs in three real-world networks. $\xi_c$ and $\xi_d$ are the medians of the corresponding distributions.  }
\end{figure*}

\section{Conclusion}
In this paper, we have discussed the node-similarity distribution of complex networks with applications to the link prediction problem. Specifically, by utilizing the generating function, we have developed a general framework to calculate the common neighbor based similarity (CNS) distribution of node sets of  an arbitrary cardinality. Furthermore, we have particularly explored the CNS distributions of various kinds of complex network models and real-world networks, and proposed an unified ring model, which unifies the  RRL, WS, NW, and ER models in terms of the CNS distributions.
Particularly, we have discovered that in the ER random network the CNS of node sets of any size follows the Poisson distribution. In the small-world network, the CNS of connected node pairs follows the bimodal distribution, while the CNS of the disconnected or all node pairs approximately follows the Poisson distribution with a long tail.  For the scale-free network, the CNS distributions of connected, disconnected and all node pairs respectively approximate the power law  with a long tail. We have also observed through empirical experiments that for most  of the real-world networks, the CNS of connected node pairs  follows the power-law distribution approximately. Whereas, for some specific real-world networks, the CNS of connected node pairs approximately obeys the exponential distribution.

In the application aspect, we have found that the performance of  link prediction is essentially determined by the similarity distributions of connected and disconnected node pairs. The farther the two distributions are apart, the better prediction accuracy will be obtained. More importantly, we have derived the theoretical solutions of the AUC and precision based on the node-similarity distributions, which allows to develop an effective way to avoid the experimental performance evaluation  and the associated large computational cost in link prediction. With these theoretical solutions, we can further investigate the fundamental limit of link predictability of complex networks.

We would like to mention that in this paper we focus only on the CNS distribution of node sets of size two in various complex networks except for the ER random network, which is completely addressed for node sets of any size. However, the CNS distribution of larger node sets in complex networks needs to be further explored, since it has potential applications in higher-order link prediction problems. When the number of sampled node sets are huge, we may classify  those node sets  based on the isomorphism type, which is the interconnection pattern of the nodes in the node set,  and then focus on the CNS distributions of the desired isomorphism types. Moreover, the application of node-similarity in the other network-related problems, such as epidemic spreading, control and game, is also worth exploration.

\appendix  
We present the definitions of some typical similarity indices  that are used in the main text, including CN, RA, AA, LP, and Katz. We denote the similarity score of node pair ($a,b$) by $S_{ab}$ and the corresponding similarity matrix by $\mathbf{S}$. $\Omega_{ab}$ is the set of common neighbors of node pair ($a,b$). $k_z$ is the degree of node $z$.  Also, $\mathbf{A}^l$ represents the $l$-th power of the adjacency matrix $\mathbf{A}$, and $(\mathbf{A}^l)_{ab}$ equals the number of paths of length $l$ between node $a$ and $b$.\\
\begin{enumerate}[i)]

\item Common neighbors (CN) index \cite{liben07}. This index quantifies the similarity of a pair of nodes as the number of common neighbors they share,
\begin{equation}
S_{ab}=|\Omega_{ab}|. \nonumber
\end{equation}
\item 	Resource Allocation (RA) index \cite{zhou09}. This index is enlightened by the resource allocation problem. Each neighbor of node $a$ gets a unit of resource from node $a$ and then equally distributes it to all its neighbors.  The amount of resource obtained by node $b$ can be taken as the similarity of nodes $a$ and $b$,
\begin{equation}
S_{ab}=\sum_{z\in \Omega_{ab}}\frac{1}{k_z}.\nonumber
\end{equation}

\item 	Adamic-Adar (AA) Index \cite{adamic03}. Similar to  CN and RA, this index is also dependent on the common neighbors,  while  each common neighbor is penalized by the inverse logarithm of its degree,
\begin{equation}
S_{ab}=\sum_{z\in \Omega_{ab}}\frac{1}{\log(k_z)}.\nonumber
\end{equation}

\item	Local Path (LP) Index \cite{lu09}. The above three indices consider only the contribution of common neighbors (paths of length 2) in link prediction. The paths of larger length between two nodes  can also reflect, to some extent, their similarity, especially when there are no common neighbors.   Besides the common neighbors, the LP index further considers the paths of length 3, which is defined as
\begin{equation}
S_{ab}=(\mathbf{A}^2)_{ab}+ \varphi (\mathbf{A}^3)_{ab},\nonumber
\end{equation}
where $ \varphi$  is a free parameter controlling the weight of paths of length 3 in the measurement of node-similarity. In the experiment, we set empirically $\varphi=0.02$, which is a proper value for the three real-world networks considered in the main text.\\

\item	Katz index \cite{katz53}. This index considers all the paths between two nodes,
\begin{equation}
S_{ab}=\sum_{l=1}^{\infty}\varphi^l(\mathbf{A}^l)_{ab},\nonumber
\end{equation}
where the free parameter $\varphi$  controls the weight of paths, and is set to be 0.01 empirically.   We often directly calculate the node-similarity matrix, $$\mathbf{S}=(\mathbf{I}-\varphi \mathbf{A})^{-1}-\mathbf{I},$$ where $\mathbf{I}$ is the identity matrix.

\end{enumerate}

Through the definition of the Katz index,  we can see that the paths of length one (links) are included, which even have the largest weight among the paths of various lengths.
 Thus, the similarity score of a connected node pair significantly changes when the link of the node pair is moved to the test set, which is undesirable in link prediction and disturbs the theoretical calculation of the AUC and precision. An approach for  re-mediation is to remove the first item in the definition of Katz to ignore the paths of length one, or simply replace $P_c(w)$ with $P_c(w+\varphi)$ in the theoretical calculation of  the AUC and precision.

\ifCLASSOPTIONcaptionsoff
  \newpage
\fi



\bibliographystyle{IEEEtran}
%








\end{document}